\documentclass{aa}  

\usepackage{graphicx}
\usepackage{txfonts}
\usepackage{lipsum}
\usepackage{subcaption}         
\usepackage{lscape}             
\usepackage{placeins}           

\usepackage{xcolor}
\definecolor{xlinkcolor}{rgb}{0, 0, 1}
\usepackage[bookmarks=true, pdfnewwindow=true, colorlinks=true, linkcolor=xlinkcolor, citecolor=xlinkcolor, filecolor=xlinkcolor, urlcolor=xlinkcolor, final=true]{hyperref}
\usepackage{bookmark}

\newcommand{\hirata}{\textup{Hirata et al. in preparation}}
\newcommand{\gu}{\textup{Gu et al. in preparation}}
\newcommand{\xa}{\textit{XRISM}}
\newcommand{\ah}{\textit{Hitomi}}
\newcommand{\ch}{\textit{Chandra}}
\newcommand{\xmm}{\textit{XMM-Newton}}
\newcommand{\suzaku}{\textit{Suzaku}}
\newcommand{\hea}{Fe~He$\alpha$}
\newcommand{\lya}{Fe~Ly$\alpha$}
\newcommand{\lyaa}{Ly$\alpha_{2}$/Ly$\alpha_{1}$}
\newcommand{\atomdb}{\textsc{AtomDB}}
\newcommand{\spex}{\textsc{spex}}
\newcommand{\chianti}{\textsc{Chianti}}

\graphicspath{{./}} 

\begin{document}

\title{Diagnosing the Fe line complex of the intracluster medium using \textit{XRISM} high-resolution spectroscopy}

\author{
K.~Fukushima\inst{1}\corrauth{kxfukushima@gmail.com} 
\href{https://orcid.org/0000-0001-8055-7113}{\includegraphics[width=8pt]{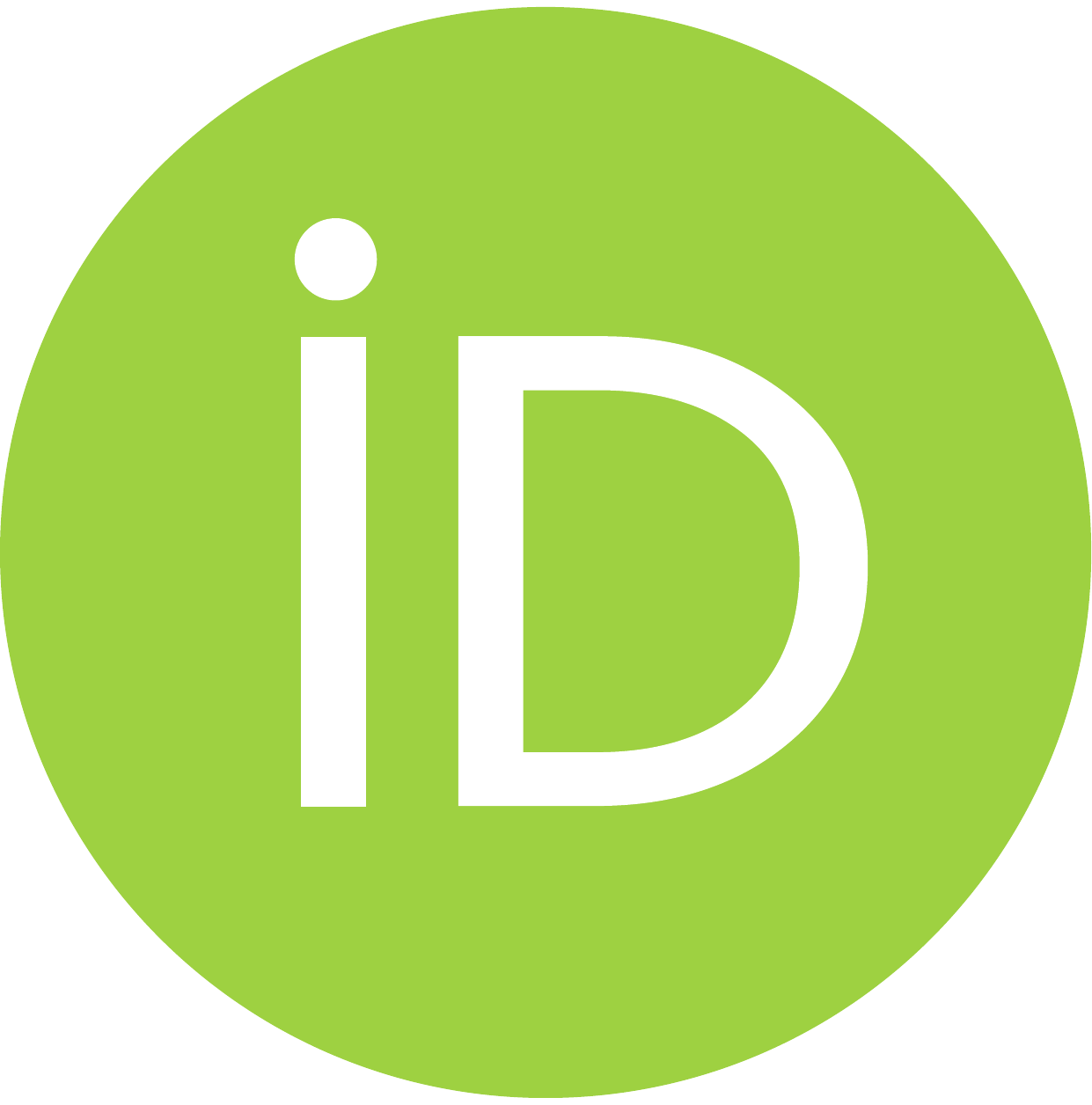}}
\and L.~Hirata\inst{2, 3}
\href{https://orcid.org/0009-0003-7156-7506}{\includegraphics[width=8pt]{logo-orcid.pdf}}
\and N.~Y.~Yamasaki\inst{3, 2}
\href{https://orcid.org/0000-0003-4885-5537}{\includegraphics[width=8pt]{logo-orcid.pdf}}
\and P.~Chakraborty\inst{4, 5}
\href{https://orcid.org/0000-0002-4469-2518}{\includegraphics[width=8pt]{logo-orcid.pdf}}
\and S.~Dupourqu\'e\inst{6}
\href{https://orcid.org/0000-0003-2715-8986}{\includegraphics[width=8pt]{logo-orcid.pdf}}
\and A.~R.~Foster\inst{4}
\href{https://orcid.org/0000-0003-3462-8886}{\includegraphics[width=8pt]{logo-orcid.pdf}}
\and Y.~Fujita\inst{7} 
\href{https://orcid.org/0000-0003-0058-9719}{\includegraphics[width=8pt]{logo-orcid.pdf}}
\and L.~Gu\inst{8} 
\href{https://orcid.org/0000-0001-9911-7038}{\includegraphics[width=8pt]{logo-orcid.pdf}}
\and C.~Kilbourne\inst{9}
\href{https://orcid.org/0000-0001-9464-4103}{\includegraphics[width=8pt]{logo-orcid.pdf}}
\and K.~Matsushita\inst{1} 
\href{https://orcid.org/0000-0003-2907-0902}{\includegraphics[width=8pt]{logo-orcid.pdf}}
\and F.~Mernier\inst{6} 
\href{https://orcid.org/0000-0002-7031-4772}{\includegraphics[width=8pt]{logo-orcid.pdf}}
\and E.~D.~Miller\inst{10} 
\href{https://orcid.org/0000-0002-3031-2326}{\includegraphics[width=8pt]{logo-orcid.pdf}}
\and K.~Nakazawa\inst{11} 
\href{https://orcid.org/0000-0003-2930-350X}{\includegraphics[width=8pt]{logo-orcid.pdf}}
\and Y.~Omiya\inst{3}
\href{https://orcid.org/0009-0009-9196-4174}{\includegraphics[width=8pt]{logo-orcid.pdf}}
\and N.~Ota\inst{12, 13}
\href{https://orcid.org/0000-0002-2784-3652}{\includegraphics[width=8pt]{logo-orcid.pdf}}
\and A.~Sarkar\inst{5, 10}
\href{https://orcid.org/0000-0002-5222-1337}{\includegraphics[width=8pt]{logo-orcid.pdf}}
\and K.~Sato\inst{14} 
\href{https://orcid.org/0000-0001-5774-1633}{\includegraphics[width=8pt]{logo-orcid.pdf}}
\and M.~Sun\inst{15}
\href{https://orcid.org/0000-0001-5880-0703}{\includegraphics[width=8pt]{logo-orcid.pdf}}
\and Y.~Uchida\inst{2, 16}
\href{https://orcid.org/0000-0002-7962-4136}{\includegraphics[width=8pt]{logo-orcid.pdf}}
\and I.~Zhuravleva\inst{17} 
\href{https://orcid.org/0000-0001-7630-8085}{\includegraphics[width=8pt]{logo-orcid.pdf}}
\and H.~Yamaguchi\inst{3, 2}
\href{https://orcid.org/0000-0002-5092-6085}{\includegraphics[width=8pt]{logo-orcid.pdf}}
}

\institute{
Department of Physics, Tokyo University of Science, 1-3 Kagurazaka, Shinjuku-ku, Tokyo 162-8601, Japan
\and Department of Physics, The University of Tokyo, 7-3-1 Hongo, Bunkyo-ku, Tokyo 113-0033, Japan
\and Institute of Space and Astronautical Science, JAXA, 3-1-1 Yoshinodai, Chuo-ku, Sagamihara, Kanagawa 252-5210, Japan
\and Center for Astrophysics $\vert$ Harvard \& Smithsonian, 60 Garden Street, Cambridge, Massachusetts, 02138
\and Department of Physics, University of Arkansas, 825 W Dickson st., Fayetteville, Arkansas 72701, USA
\and Univ Toulouse, CNES, CNRS, IRAP, Toulouse, France
\and Department of Physics, Tokyo Metropolitan University, 1-1 Minami-Osawa, Hachioji, Tokyo 192-0397, Japan
\and SRON Space Research Organisation Netherlands, Niels Bohrweg 4, 2333 CA Leiden, The Netherlands
\and NASA/Goddard Space Flight Center, Greenbelt, Maryland 20771, USA
\and Kavli Institute for Astrophysics and Space Research, Massachusetts Institute of Technology, Massachusetts 02139, USA
\and Kobayashi-Maskawa Institute for the Origin of Particles and the Universe, Nagoya University, Furocho, Chikusa-ku, Nagoya, Aichi 464-8602, Japan
\and Department of Physics, Nara Women's University, Kitauoyanishi-machi, Nara, Nara 630-8506, Japan
\and Argelander-Institut f\"ur Astronomie (AIfA), Universit\"at Bonn, Auf dem H\"ugel 71, 53121 Bonn, Germany
\and Department of Astrophysics and Atmospheric Sciences, Kyoto Sangyo University, Motoyama, Kamigamo, Kita-ku, Kyoto, Kyoto 603-8555, Japan
\and Department of Physics and Astronomy, University of Alabama in Huntsville, Huntsville, Alabama, USA
\and Research Center for Space System Innovation, Research Institute for Science and Technology, Tokyo University of Science, 2641 Yamazaki, Noda, Chiba 278-8510, Japan
\and Department of Astronomy and Astrophysics, University of Chicago, Chicago, Illinois 60637, USA
}



\abstract
{
With the advent of high-resolution X-ray spectroscopy by \xa{}/Resolve,
achieving an energy resolution of 5\,eV at 6\,keV, individual fine structures within the Fe-K complex,
such as the He-like \ion{Fe}{xxv} triplet and the H-like \ion{Fe}{xxvi} doublet, can now be resolved.
This enables precise measurements of velocity fields, temperature,
and elemental abundances in the intracluster medium (ICM),
which has long been believed to be a collisional ionisation equilibrium (CIE) plasma.
}
{
We aim to test the validity of the CIE framework in the ICM by performing line diagnostics
based mainly on resolved Fe-K emission lines.
}
{
We analyse Resolve full-array spectra of 17 galaxy clusters.
Prominent Fe-K line components (the \ion{Fe}{xxv} $w$, $x$, $y$, $z$, and \ion{Fe}{xxvi} Ly$\alpha_{1, 2}$ lines)
are removed from plasma emission models and instead fitted with Gaussian profiles,
enabling direct measurements of line fluxes without relying on synthetic spectral models.
}
{
Some cool-core systems show $w/z$ ratios lower than predicted by about 20 per cent,
and a broader $w$ than $z$, consistent with resonant scattering effects.
The $y/x$ ratios exhibit marginal deviations from model predictions for some objects,
suggesting possible origins of the cascade process due to electron recombination
and contribution from low-ionised Fe.
The Fe~\lyaa{} ratios are globally close to the expected value of about 0.5,
and the samples with good photon statistics prefer 0.55.
This subtle excess is consistent with an unresolved contribution to Ly$\alpha_{2}$
from a magnetic-dipole transition, which is absent from one of the atomic codes considered here.
More interestingly, systems at around 7\,keV preferentially exhibit \lyaa{} ratios above 0.55.
Although the statistical significance of this trend is limited, it suggests that resolved \lya{} spectroscopy
may provide a sensitive probe of additional atomic processes beyond collisional excitation,
including dielectronic and radiative recombination and polarisation effects.
}
{
While the temperature dependence of Fe~\lyaa{} remains tentative,
resolved \hea{} and \lya{} spectroscopy provides new diagnostics of subtle atomic processes in the ICM
that conventional spectral modelling may not fully capture.
Our results motivate a reassessment of the simple CIE description of the ICM in the \xa{} era
through cross-disciplinary efforts in astrophysics and atomic physics.
}


\maketitle
\nolinenumbers

\section{Introduction \label{sec:intro}}

X-ray spectroscopy of galaxy clusters has long focused on Fe K-shell emission lines,
particularly the He-like \ion{Fe}{xxv} line at 6.7\,keV (\hea{})
and the H-like \ion{Fe}{xxvi} line at 7.0\,keV (\lya{}),
since their first detections in the intracluster medium (ICM, \citealt{Mitchell76, Serlemitsos77}).
Legacy missions such as \ch{}, \xmm{}, and \suzaku{} confirmed
the thermodynamics \citep[e.g.,][]{Ota07, Sato11, Werner16}
and the enrichment processes of the ICM \citep[e.g.,][]{Matsushita07, Panagoulia15, Mernier18a}.
These studies established Fe lines as principal tools
for probing the thermodynamic state and chemical enrichment history of the ICM.
However, their limited energy resolution prevented sufficient separation of the \hea{} resonance,
intercombination, forbidden lines, and the doublet structure of \lya{}.
Detecting Doppler shifts and line broadening was also challenging \citep[e.g.,][for reviews]{Gu18, Sanders23}.
Thus, key physical processes, such as turbulent energy dissipation,
feedback efficiency, and multi-temperature structure, remained hard or strongly degenerate.

A major breakthrough in high-resolution ICM spectroscopy
over the past half-century was achieved by the short-lived \ah{} satellite \citep[][]{Takahashi18}.
Observations of the Perseus cluster core demonstrated the diagnostic power of resolving
the fine structure of Fe-K emission lines at an energy resolution of 5\,eV for the first time.
By sufficiently resolving the \hea{} triplet, \ah{} enabled direct and non-degenerate measurements
of the ICM velocity field, revealing a subsonic turbulent velocity of about 160\,km\,s$^{-1}$
and a correspondingly small non-thermal pressure fraction \citep[][]{Hitomi18a}.
In the same framework, suppression of the \hea{} resonance line relative to the forbidden line,
due to a resonant scattering effect, gave an independent and consistent constraint
on the velocity dispersion \citep[][]{Hitomi18b}.
Furthermore, the \lya{} line, resolved with sufficient precision,
offered an additional probe of the temperature structure of the core \citep[][]{Hitomi18c}.
These results collectively demonstrated that resolving the Fe-K line complex
presents a unified and powerful diagnostic of the thermodynamic and dynamical state of the ICM.

The international X-ray mission \xa{} led by Japan \citep[][]{Tashiro25}, equipped with the Resolve instrument,
is designed to extend the high-resolution spectroscopic capability pioneered by \ah{}.
The power of Resolve to separate both the \hea{} and \lya{} complexes enables us
to conduct systematic studies of ICM thermodynamics beyond the Perseus cluster.
Recent \xa{} observations have measured velocity dispersions of the central ICM in multiple systems,
typically at levels of 100--300\,km\,s$^{-1}$
\citep[][]{Fujita25, Rose25, XRISM25h, XRISM25c, XRISM25d, XRISM25a, Aihara26, Sarkar26a},
at unprecedented precision typically lower than tens of kilometres per second.
In addition, hints of resonant scattering have been reported in a strong cool-core system,
providing another probe of velocity structure
(e.g., PKS~0745$-$191, \citealt{Tanaka26}; A2199, \citealt{Suda26}).
Complex temperature structures have also been characterised in several clusters
(e.g., A2029, \citealt{Sarkar25a}; Centaurus, \citealt{Kondo26}).
These results use line-resolved diagnostics that depend less on global spectral modelling.
Recent studies have also explored the Fe-K complex as a probe
of departures from collisional ionisation equilibrium (CIE).
For example, \citet{Churazov26b} propose that the \hea{} triplet would serve as a powerful probe
for identifying the non-equilibrium ionisation (NEI) process in the ICM.

As the \xa{} results accumulate, they are beginning to reveal subtle but systematic deviations
from the conventional picture of the ICM as a simple CIE plasma.
For example, excess emission in the intercombination $y$ line at the level of about 50--60 per cent
has been reported in several clusters
\citep[e.g.,][]{Fujita25, XRISM25c, Heinrich26a, Sarkar26a, Suda26},
as well as the Coma cluster exhibits an anomalous \lya{} doublet
with an approximately unity intensity ratio \citep[][]{XRISM25d},
in contrast to the canonical one-to-two ratio expected from atomic transition probabilities
under optically thin and isotropic conditions.
Such inconsistencies imply that even the most basic assumptions of ICM diagnostics,
particularly the analysis of Fe-K line ratios within the CIE framework,
would necessitate reassessment.
Given that these line ratios are essential for assessing temperatures and/or elemental abundances,
any systematic deviation can bias our insight into cluster thermodynamics and chemical evolution.

Motivated by this emerging tension, we focus on individual emission lines arising from the fine structure
of ionised Fe species across multiple galaxy clusters.
By measuring their intensities using an approach that does not rely on
synthetic CIE spectral models (e.g., APEC; \citealt{Smith01}),
we aim to establish line-based diagnostics that can directly test the validity of the ``textbook'' plasma paradigm.
This paper is organised as follows.
In Section\,\ref{sec:obs}, we summarise our \xa{} sample of galaxy clusters and data reduction.
In Section\,\ref{sec:analysis}, prescriptions of our spectral analysis are described.
We interpret and discuss the results, providing possible insights into the plasma state
in the ICM in Section\,\ref{sec:result}.
In this article, we adopt a standard $\Lambda$CDM cosmology
with $H_0=70$\,km\,s$^{-1}$\,Mpc$^{-1}$, $\Omega_m=0.3$, and $\Omega_{\Lambda}=0.7$.
Statistical uncertainties are given at a confidence level of 1$\sigma$ unless otherwise stated.

\section{Observations and data reduction \label{sec:obs}}

\begin{table*}
\centering
\caption{The \xa{} observations analysed in this work. \label{tab:obs}}

\begin{tabular}{@{}l@{\:\:}c@{\:\:}c@{\:\:}c@{\:\:}c@{\:\:}c}\\ \hline\hline
Object\tablefootmark{a} & ObsID\tablefootmark{b} & $\alpha$, $\delta$\tablefootmark{c} & $N_\textup{H}$\tablefootmark{d} & redshift\tablefootmark{e} & Exposure\tablefootmark{f} \\
 & & (deg) & ($10^{20}$\,cm$^{-2}$) & & (ks) \\ \hline
Virgo & 300014010 & $187.704$, $12.390$ & $1.3$ & $4.3 \times 10^{-3}$ & $121.7$ \\
Centaurus & 000138000 & $192.196$, $-41.300$ & $7.8$ & $9.9 \times 10^{-3}$ & $298.4$ \\
Centaurus West & 201031010 & $192.149$, $-41.334$ & = & = & $296.8$ \\
Perseus & 00015[4, 5]000, 1010[09, 10, 11, 12]010 & $49.950$, $41.512$ & $14$ & $1.8 \times 10^{-2}$ & $364.7$ \\
Coma & 300073010 & $194.944$, $27.947$ & $0.93$ & $2.4 \times 10^{-2}$ & $402.1$ \\
Coma South & 3000740[1, 2]0 & $194.941$, $27.847$ & = & = & $171.0$ \\
Coma North & 201114010 & $194.941$, $28.049$ & = & = & $144.8$ \\
Ophiuchus & 201006010 & $258.114$, $-23.369$ & $19$ & $2.8 \times 10^{-2}$ & $217.1$ \\
Ophiuchus South-West& 201117010 & $258.129$, $-23.393$ & = & = & $116.6$ \\
A2199 & 201089010 & $247.159$, $39.551$ & $0.81$ & $3.0 \times 10^{-2}$ & $262.6$ \\
A3571 & 201095010 & $206.866$, $-32.853$ & $3.9$ & $3.9 \times 10^{-2}$ & $192.1$ \\
A3571 North & 201024010 & $206.866$, $-32.837$ & = & = & $137.4$ \\
A3571 South & 201023010 & $206.866$, $-32.888$ & = & = & $174.6$ \\
A3571 East & 201025010 & $206.928$, $-32.855$ & = & = & $72.4$ \\
A3571 South-East & 202076010 & $206.926$, $-32.905$ & = & = & $69.3$ \\
A2319 & 00010[2, 3]000 & $290.300$, $43.924$ & $8.4$ & $5.5 \times 10^{-2}$ & $138.0$ \\
A3667 & 201051010 & $303.265$, $-56.886$ & $4.3$ & $5.5 \times 10^{-2}$ & $298.9$ \\
A3667 (inside cold front) & 201050010 & $303.174$, $-56.848$ & = & = & $116.6$ \\
Hydra~A & 201070010 & $139.523$, $-12.095$ & $4.0$ & $5.5 \times 10^{-2}$ & $116.3$ \\
Cygnus~A & 201120010 & $299.867$, $40.734$ & $4.0$ & $5.6  \times 10^{-2}$ & $179.4$ \\
A1795 & 201087010 & $207.218$, $26.592$ & $1.0$ & $6.1 \times 10^{-2}$ & $225.7$ \\
A1795 North & 201088010 & $207.218$, $26.643$ & = & = & $113.4$ \\
A2029 & 0001[49, 51]000, 201042010 & $227.733$, $5.745$ & $3.0$ & $7.8 \times 10^{-2}$ & $164.0$ \\
A2029 North1 & 000150000 & $227.764$, $5.785$ & = & = & $106.2$ \\
A2029 North2 & 300053010 & $227.794$, $5.825$ & = & = & $366.2$ \\
A478 & 202059010 & $63.356$, $10.465$ & $15$ & $8.6 \times 10^{-2}$ & $123.3$ \\
PKS~0745$-$191 & 000112000 & $116.881$, $-19.295$ & $41$ & $0.10$ & $21.3$ \\
A1413 & 201049010 & $178.827$, $23.410$ & $1.8$ & $0.14$ & $107.8$ \\
A1914 & 201093010 & $216.500$, $37.826$ & $0.99$ & $0.16$ & $111.8$ \\ \hline
\end{tabular}

\tablefoot{
\tablefoottext{a}{Object name.}
\tablefoottext{b}{Observation ID in the \xa{} archive.}
\tablefoottext{c}{Pointing positions of each observation in the J2000.0 equatorial coordinates.}
\tablefoottext{d}{Galactic absorption retrieved from \citet{HI4PI16}.}
\tablefoottext{e}{Redshift of BCGs in each cluster derived from NASA/IPAC Extragalactic Database (NED).}
\tablefoottext{f}{Cleaned exposure time of Resolve.}
}

\end{table*}

In this work, we analysed observational data from 17 galaxy clusters using \xa{}.
Table~\ref{tab:obs} lists the target clusters and pointings analysed, along with their properties.
We focused exclusively on data from the Resolve instrument \citep[][]{Ishisaki25, Kelley25},
which were processed by the \xa{} Science Data Center (SDC) using the version 3 pipeline software
\footnote{\url{https://data.darts.isas.jaxa.jp/pub/xrism/data/00readme_data.txt}}.
Following the procedures described for the Centaurus cluster \citep[][]{XRISM25a},
we applied further screening recommended by the \xa{} team,
using the calibration database (\texttt{CALDB}) version 11,
including the Resolve response parameters (\texttt{20190101v007}).
Table~\ref{tab:obs} lists the cleaned exposure times for each dataset.
Gain correction for the Resolve data was performed with standard gain-tracking data
from the $^{55}$Fe calibration sources (see \citealt{Porter25, Sawada25},
and the Resolve energy scale quality reports for each observation provided by SDC
\footnote{\url{https://heasarc.gsfc.nasa.gov/FTP/xrism/postlaunch/gainreports/}}).
We note that the PKS~0745$-$191 and A2319 datasets, obtained during the initial commissioning phase,
are subject to specific caveats regarding gain reconstruction.
Following the prescriptions provided in the literature
(\citealt{XRISM25h} for A2319; \citealt{Tanaka26} for PKS~0745$-$191),
the data quality for these targets was recovered.

The Resolve spectra were constructed by integrating high-primary events (\texttt{ITYPE=0}) from all pixel channels,
achieving a full width at half maximum (FWHM) of 5\,eV at 6\,keV \citep[][]{Ishisaki25} on the entire field of view (FoV).
Channel 12 is installed as a calibration pixel irradiated by the $^{55}$Fe source and not exposed to celestial emissions.
We also removed channel 27 due to anomalous gain jumps,
except for the data from the Centaurus centre (see \citealt{XRISM25a}).
During observations of A2029 (\texttt{201042010}) and the Centaurus south-west region (\texttt{201031010}),
channel 7 exhibited an abnormal increase in noise levels and significant degradation in energy resolution
\footnote{\url{https://xrism.isas.jaxa.jp/research/observers/operation_log/Resolve/index.html}};
thus, this channel was also excluded from both observations.
We generated redistribution matrix files (RMFs) using the \texttt{rslmkrmf} task with \texttt{whichrmf=X},
accounting for the Gaussian core, exponential tail, escape peaks, Si~K$\alpha$ emission line,
and electron loss continuum components for each line spread function.
Ancillary response files (ARFs) were calculated via ray-tracing simulations
using the \texttt{xrtraytrace} and \texttt{xaxmaarfgen} procedures,
with the ray-traced photon distribution based on 2--8\,keV X-ray flux images from \ch{} or \xmm{}.

\section{Measurements of line flux ratios \label{sec:analysis}}

\begin{table}
\centering
\caption{Major lines around the \hea{} and \lya{} complexes. \label{tab:line}}

\begin{tabular}{lcccc}\\ \hline\hline
Label\tablefootmark{a} & Energy & Ion & Configuration \\
 & (keV) & \\ \hline
$z$ \tablefootmark{b} & $6.637$ & \ion{Fe}{xxv} & 1s$^{2}$ $^1$S$_0$--1s2s $^3$S$_1$ \\
$j$ & $6.645$ & \ion{Fe}{xxiv} & 1s$^{2}$2p $^2$P$_{3/2}$--1s2p$^2$ $^2$D$_{5/2}$ \\
$q$ & $6.662$ & \ion{Fe}{xxiv} & 1s$^{2}$2s $^2$S$_{1/2}$--1s2s($^3$S)2p $^2$P$_{3/2}$ \\
$y$ \tablefootmark{b}  & $6.668$ & \ion{Fe}{xxv} & 1s$^{2}$ $^1$S$_0$--1s2p $^3$P$_1$ \\
$x$ \tablefootmark{b}  & $6.682$ & \ion{Fe}{xxv} & 1s$^{2}$ $^1$S$_0$--1s2p $^3$P$_2$ \\
$w$ \tablefootmark{b}  & $6.700$ & \ion{Fe}{xxv}  & 1s$^{2}$ $^1$S$_0$--1s2p $^1$P$_1$ \\
$J$ & $6.918$ & \ion{Fe}{xxv} & 1s2p $^1$P$_1$--2p$^2$ $^1$D$_2$ \\
Ly$\alpha_2$ \tablefootmark{b}  & $6.952$ & \ion{Fe}{xxvi} & 1s $^2$S$_{1/2}$--2p $^2$P$_{1/2}$ \\
M1 \tablefootmark{c} & $6.952$ & \ion{Fe}{xxvi} & 1s $^2$S$_{1/2}$--2s $^2$S$_{1/2}$ \\
Ly$\alpha_1$ \tablefootmark{b}  & $6.973$ & \ion{Fe}{xxvi} & 1s $^2$S$_{1/2}$--2p $^2$P$_{3/2}$ \\ \hline
\end{tabular}

\tablefoot{
\tablefoottext{a}{Labels for lines follow the notation of \citet{Gabriel72} and \citet{Kurihara25}.}
\tablefoottext{b}{Removed from the atomic code in our analysis.}
\tablefoottext{c}{This transition is not included in \atomdb{} version 3.1.3.}
}

\end{table}

\begin{figure}

\centering
\includegraphics[width=\columnwidth]{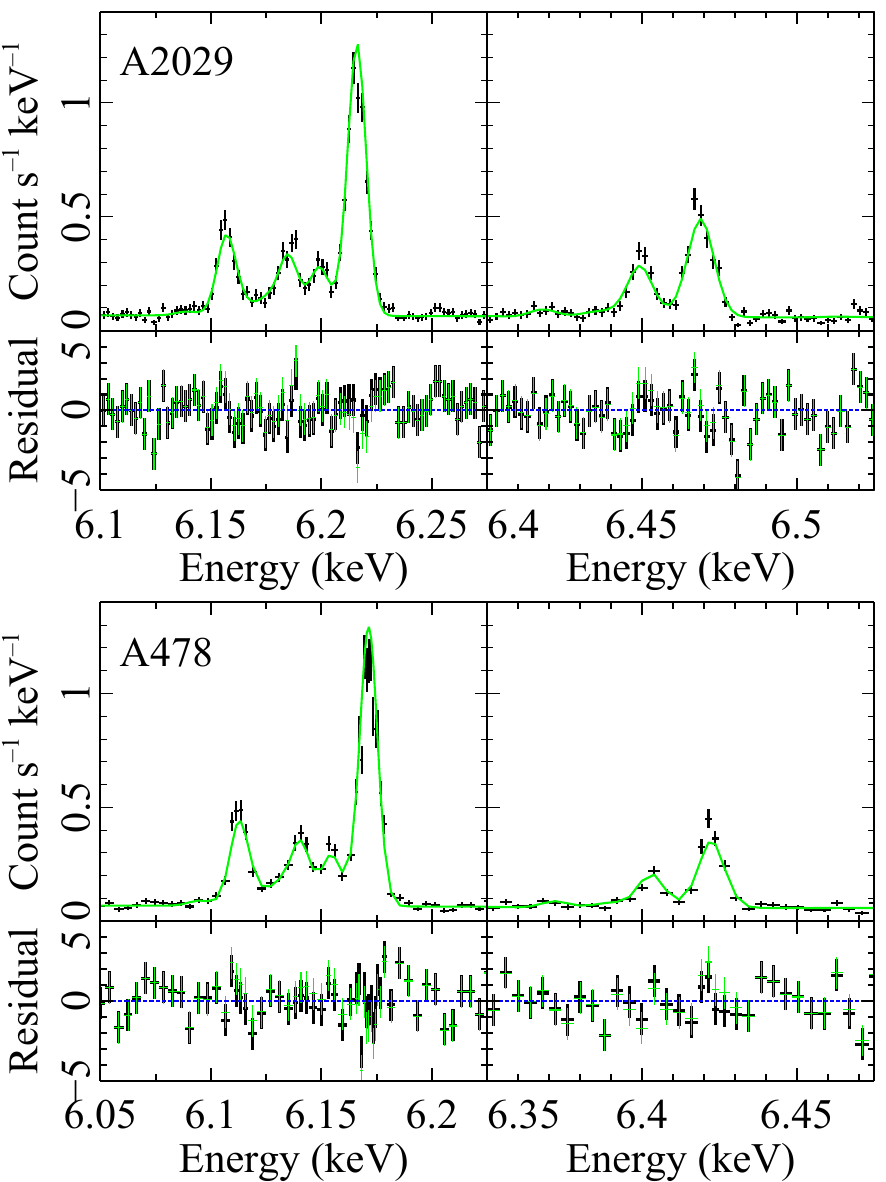}
\caption{Representative spectral fitting results for A2029 and A478.
Solid lines indicate the best-fitting isothermal CIE models.
In the bottom panels, residuals are provided in terms of $(\textup{data} - \textup{model})/\sqrt{\textup{model}}$
for the CIE model (thin) and the benchmark model described in Section~\ref{sec:analysis} (thick).
\label{fig:spec}}

\end{figure}

\begin{figure}

\centering
\includegraphics[width=\columnwidth]{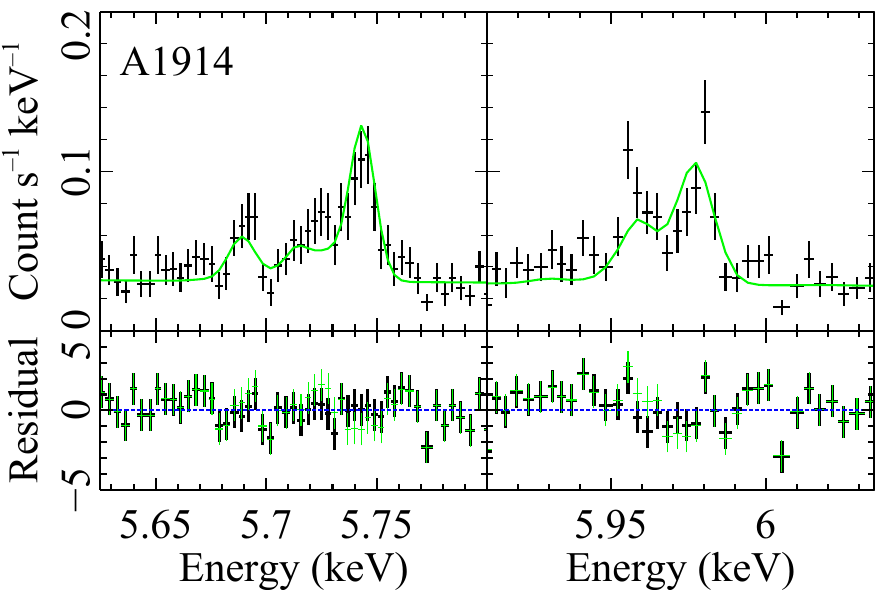}
\caption{Same as Fig.~\ref{fig:spec} for A1914 that is affected by a strong multi-velocity contamination.
\label{fig:specv}}

\end{figure}

We analyse the Resolve spectra extracted from the entire FoV in the 1.8--12.0\,keV band
using \textsc{xspec} version 12.15.0 \citep[][]{Arnaud96} and \atomdb{} version 3.1.3 \citep[][]{Smith01, Foster12}.
The non-X-ray background (NXB) is modelled using the standard \xa{} NXB spectral components,
consisting of a power-law continuum and Gaussian components representing instrumental lines
\footnote{\url{https://heasarc.gsfc.nasa.gov/docs/xrism/analysis/nxb/nxb_spectral_models.html}}.
For M87, Perseus, Hydra~A, and Cygnus~A, we additionally include the active galactic nucleus component
following previous analyses of each target \citep[][]{Rose25, Majumder26a, XRISM26b, XRISM26c}.
No additional astrophysical background components, including the cosmic X-ray background,
are included because the ICM emission dominates the Resolve spectra by more than an order of magnitude
within the field of view \citep[e.g.,][]{XRISM25c, XRISM25e, XRISM25a, Fukushima26}.
The spectra are fitted by minimising the $C$ statistic \citep[][]{Cash79},
which provides unbiased parameter estimates and uncertainties for Poisson-distributed data \citep[][]{Kaastra17}.

To reproduce the ICM continuum emission, we employ the absorbed CIE plasma model
with the \texttt{bapec} component in \textsc{xspec}.
The Galactic hydrogen column density, $N_\mathrm{H}$,
is fixed to the values obtained from \citet{HI4PI16}, as listed in Table~\ref{tab:obs}.
As described in Section~\ref{sec:intro}, our primary goal
is to measure flux ratios of emission lines from highly ionised Fe.
Such line-ratio diagnostics are widely used to probe the physical properties of cosmic plasmas,
including supernova remnants, stellar coronae and winds, and the ICM
\citep[e.g.,][]{Watanabe24, Kurihara25, Mochizuki25a, Sarkar25a, XRISM25b, Kondo26}.

Table~\ref{tab:line} lists the principal Fe-K lines and nearby satellite transitions.
To measure the fluxes of individual Fe lines, we remove the six strongest transitions
($z$, $y$, $x$, $w$, Ly$\alpha_2$, and Ly$\alpha_1$) from the \texttt{bapec} model
and replace them with \texttt{zgauss} components
\footnote{We note that this modelling is an approximation.
The line profile can deviate from a simple Gaussian form due to some emission processes
(e.g., an \textit{M-shape} profile of the $w$ line by resonant scattering,
see \citet{Churazov10b} and Section~\ref{subsec:zw}).}.
The centroid energies in the rest frame are fixed to the values listed in Table~\ref{tab:line}.
Line widths ($\sigma_\mathrm{Fe}$) and redshifts are linked
among the six Gaussian components and are allowed to vary freely.
Using this model, we simultaneously determine the ICM temperature ($kT$) and the fluxes of the six Fe lines.
The $kT$ value is primarily constrained by the shape of the continuum.
We note that the measured flux ratios among lines from the same ions (e.g., \hea{}~$y$/$x$, Fe~\lyaa{})
are not affected significantly by multiple-temperature modelling, whereas those between distinct ions
(e.g., Ly$\alpha_1$/$z$) fluctuate, indicating the difference between \textit{local-band} temperature
around the Fe~K complex and $kT$ determined from the continuum
\footnote{Indeed, a few systems with strong intermediate-mass-element lines, such as Virgo, Centaurus, and A2029,
are well suited to investigate multi-temperature structures \citep[e.g.,][]{Sarkar25a, Kondo26, Simionescu26}.
For simplicity, however, we fit each spectrum with a single-temperature model in this work.}.
Figure~\ref{fig:spec} shows representative spectra and best-fitting models for A2029 and A478.
The line-based model substantially reduces residuals around the Fe-K complex
compared with the standard isothermal CIE model.

One limitation of this approach is that distorted line profiles may bias the inferred line fluxes
because the model assumes Gaussian line shapes.
Even after applying the benchmark model, A1914 exhibits residuals at 5.67\,keV and 5.94\,keV,
for instance, near the Fe line complexes (Fig.~\ref{fig:specv}).
These residuals may indicate the presence of an additional velocity component
blueshifted by more than 1000\,km\,s$^{-1}$.
Indeed, \citet{Heinrich25} showed that a multi-velocity model gives a significantly better description
of the A1914 spectrum than a simple line-based model resembling ours.
A similar interpretation has been proposed for A2034, another major merger system \citep[][]{Heinrich26a}.
These studies indicate that multiple velocity components, and possibly additional NEI plasma components,
can distort the observed \hea{} and \lya{} profiles.
Other disturbed systems in our sample (A2319, Coma, A3667) show more moderate velocity gradients
in the Resolve FoV up to 300\,km\,s$^{-1}$ \citep[][]{XRISM25h, XRISM25d, Omiya26a}.
Such an exceptionally large velocity gradient complicates the interpretation of line flux ratios;
therefore, we exclude A1914 from the subsequent discussion.


\begin{table*}
\centering
\caption{Observed spectral parameters and line ratios of \hea{} and \lya{}. \label{tab:ratio}}

\begin{tabular}{@{}l@{\:\:}c@{\:\:}c@{\:\:}c@{\:\:}c@{\:\:}c@{\:\:}c@{\:\:}c@{\:\:}c}\\ \hline\hline
Object & $kT$ & $\sigma_\textup{Fe}$\tablefootmark{a} & \lya{}\tablefootmark{b} & $y$/$z$ & $x$/$z$ & $w$/$z$ & Ly$\alpha_{2}$/$z$ & Ly$\alpha_{1}$/$z$ \\
 & (keV) & (eV) & (count) & & & & & \\ \hline
Virgo & $2.33 \pm 0.06$ & $2.8 \pm 0.3$ & $123$ & $0.7 \pm 0.3$ & $0.6 \pm 0.2$ & $2.4 \pm 0.6$ & $< 0.05$ & $< 0.1$ \\
Centaurus & $2.35 \pm 0.02$ & $2.8 \pm 0.1$ & $291$ & $0.63 \pm 0.06$ & $0.71 \pm 0.06$ & $2.1 \pm 0.2$ & $0.05 \pm 0.02$ & $0.09 \pm 0.02$ \\
CentaurusW & $2.96 \pm 0.04$ & $4.0 \pm 0.2$ & $165$ & $0.6 \pm 0.1$ & $0.66 \pm 0.09$ & $2.6 \pm 0.2$ & $0.06 \pm 0.03$ & $0.13 \pm 0.04$ \\
Hydra A & $3.5 \pm 0.2$ & $4.0 \pm 0.2$ & $243$ & $0.7 \pm 0.1$ & $0.55 \pm 0.09$ & $2.3 \pm 0.3$ & $0.13 \pm 0.05$ & $0.14 \pm 0.05$ \\
Perseus & $3.67 \pm 0.01$ & $4.06 \pm 0.04$ & $8903$ & $0.66 \pm 0.02$ & $0.61 \pm 0.02$ & $2.47 \pm 0.06$ & $0.17 \pm 0.01$ & $0.30 \pm 0.01$ \\
A2199 & $3.72 \pm 0.03$ & $2.81 \pm 0.09$ & $761$ & $0.77 \pm 0.07$ & $0.68 \pm 0.06$ & $2.7 \pm 0.2$ & $0.17 \pm 0.03$ & $0.32 \pm 0.04$ \\
A3667CFin & $4.6 \pm 0.1$ & $6.7 \pm 0.7$ & $153$ & $0.6 \pm 0.2$ & $0.5 \pm 0.2$ & $3.0 \pm 0.6$ & $0.3 \pm 0.1$ & $0.6 \pm 0.2$ \\
A1795 & $4.84 \pm 0.05$ & $3.30 \pm 0.08$ & $1442$ & $0.79 \pm 0.06$ & $0.55 \pm 0.05$ & $3.1 \pm 0.2$ & $0.31 \pm 0.03$ & $0.57 \pm 0.05$ \\
Cygnus A & $5.1 \pm 0.1$ & $6.0 \pm 0.2$ & $2164$ & $0.56 \pm 0.09$ & $0.6 \pm 0.1$ & $3.2 \pm 0.3$ & $0.38 \pm 0.06$ & $0.63 \pm 0.07$ \\
A1795N & $5.5 \pm 0.2$ & $3.1 \pm 0.2$ & $208$ & $0.6 \pm 0.2$ & $0.8 \pm 0.2$ & $2.9 \pm 0.5$ & $0.4 \pm 0.1$ & $0.8 \pm 0.2$ \\
A478 & $5.62 \pm 0.06$ & $3.8 \pm 0.1$ & $1198$ & $0.66 \pm 0.06$ & $0.57 \pm 0.06$ & $2.7 \pm 0.2$ & $0.32 \pm 0.04$ & $0.79 \pm 0.06$ \\
PKS0745 & $5.7 \pm 0.2$ & $4.2 \pm 0.3$ & $180$ & $0.5 \pm 0.2$ & $0.6 \pm 0.2$ & $2.4 \pm 0.4$ & $0.4 \pm 0.1$ & $0.7 \pm 0.2$ \\
A3667 & $5.8 \pm 0.2$ & $4.8 \pm 0.4$ & $168$ & $0.4 \pm 0.2$ & $0.4 \pm 0.1$ & $2.2 \pm 0.4$ & $0.3 \pm 0.1$ & $0.6 \pm 0.2$ \\
A3571 & $6.3 \pm 0.1$ & $3.4 \pm 0.1$ & $1230$ & $0.72 \pm 0.08$ & $0.55 \pm 0.07$ & $3.1 \pm 0.2$ & $0.80 \pm 0.08$ & $1.3 \pm 0.1$ \\
A2029 & $6.50 \pm 0.07$ & $4.39 \pm 0.05$ & $2022$ & $0.75 \pm 0.06$ & $0.54 \pm 0.05$ & $3.0 \pm 0.2$ & $0.69 \pm 0.05$ & $1.18 \pm 0.08$ \\
A3571SE & $6.5 \pm 0.4$ & $3.9 \pm 0.4$ & $93$ & $0.8 \pm 0.6$ & $< 0.7$ & $5 \pm 2$ & $1.2 \pm 0.7$ & $2.7 \pm 1.4$ \\
ComaS & $6.9 \pm 0.2$ & $5.1 \pm 0.5$ & $257$ & $1.6 \pm 0.8$ & $0.9 \pm 0.6$ & $6 \pm 2$ & $1.5 \pm 0.8$ & $2.8 \pm 1.2$ \\
A3571N & $7.0 \pm 0.1$ & $3.1 \pm 0.1$ & $736$ & $0.59 \pm 0.08$ & $0.69 \pm 0.09$ & $3.2 \pm 0.3$ & $0.57 \pm 0.08$ & $1.0 \pm 0.1$ \\
A1413 & $7.1 \pm 0.3$ & $4.6 \pm 0.3$ & $316$ & $0.5 \pm 0.2$ & $0.2 \pm 0.1$ & $3.2 \pm 0.6$ & $0.7 \pm 0.2$ & $1.2 \pm 0.2$ \\ 
A2029N1 & $7.1 \pm 0.3$ & $3.8 \pm 0.4$ & $247$ & $0.8 \pm 0.3$ & $0.4 \pm 0.2$ & $3.7 \pm 0.9$ & $1.3 \pm 0.4$ & $1.7 \pm 0.5$ \\
Ophiuchus & $7.19 \pm 0.08$ & $4.49 \pm 0.08$ & $3808$ & $0.72 \pm 0.04$ & $0.51 \pm 0.04$ & $2.9 \pm 0.1$ & $0.69 \pm 0.04$ & $1.31 \pm 0.07$ \\
A3571S & $7.4 \pm 0.1$ & $3.7 \pm 0.2$ & $781$ & $0.7 \pm 0.1$ & $0.6 \pm 0.1$ & $3.2 \pm 0.3$ & $0.9 \pm 0.1$ & $1.4 \pm 0.2$ \\
A3571E & $7.7 \pm 0.3$ & $4.9 \pm 0.5$ & $150$ & $0.7 \pm 0.3$ & $0.4 \pm 0.2$ & $3.4 \pm 0.9$ & $0.9 \pm 0.3$ & $1.4 \pm 0.4$ \\
Coma & $7.8 \pm 0.2$ & $5.4 \pm 0.3$ & $1005$ & $0.5 \pm 0.1$ & $0.7 \pm 0.2$ & $3.1 \pm 0.4$ & $1.2 \pm 0.2$ & $1.9 \pm 0.3$ \\
ComaN & $8.2 \pm 0.5$ & $4.1 \pm 0.6$ & $173$ & $1.5 \pm 1.1$ & $1.6 \pm 1.2$ & $6 \pm 3$ & $2.2 \pm 1.5$ & $4 \pm 3$ \\
A2319 & $8.4 \pm 0.2$ & $6.2 \pm 0.3$ & $259$ & $0.8 \pm 0.2$ & $0.4 \pm 0.2$ & $3.3 \pm 0.6$ & $1.2 \pm 0.3$ & $2.6 \pm 0.5$ \\
A2029N2 & $8.4 \pm 0.6$ & $4.7 \pm 0.6$ & $405$ & $0.5 \pm 0.4$ & $0.9 \pm 0.5$ & $4.1 \pm 1.7$ & $1.8 \pm 0.8$ & $2.2 \pm 1.0$ \\
OphiuchusSW & $8.5 \pm 0.2$ & $4.7 \pm 0.2$ & $1453$ & $0.78 \pm 0.09$ & $0.46 \pm 0.07$ & $3.1 \pm 0.3$ & $0.83 \pm 0.09$ & $1.6 \pm 0.2$ \\
A1914\tablefootmark{c} & $9.4 \pm 0.3$ & $5.4 \pm 0.5$ & $489$ & $0.7 \pm 0.3$ & $1.1 \pm 0.3$ & $2.1 \pm 0.5$ & $1.7 \pm 0.5$ & $2.0 \pm 0.5$ \\ \hline
\end{tabular}

\tablefoot{
\tablefoottext{a}{Shared among all satellite lines of \hea{} and \lya{}.}
\tablefoottext{b}{Raw photon count in the \lya{} doublet (6.920--7.000\,keV range in the rest frame).}
\tablefoottext{c}{Not included in the discussion and interpretation parts
due to a strong distortion of line profiles from a multi-velocity structure.}
}

\end{table*}

\section{Results and discussion \label{sec:result}}

\begin{figure*}

\centering
\includegraphics[width=\textwidth]{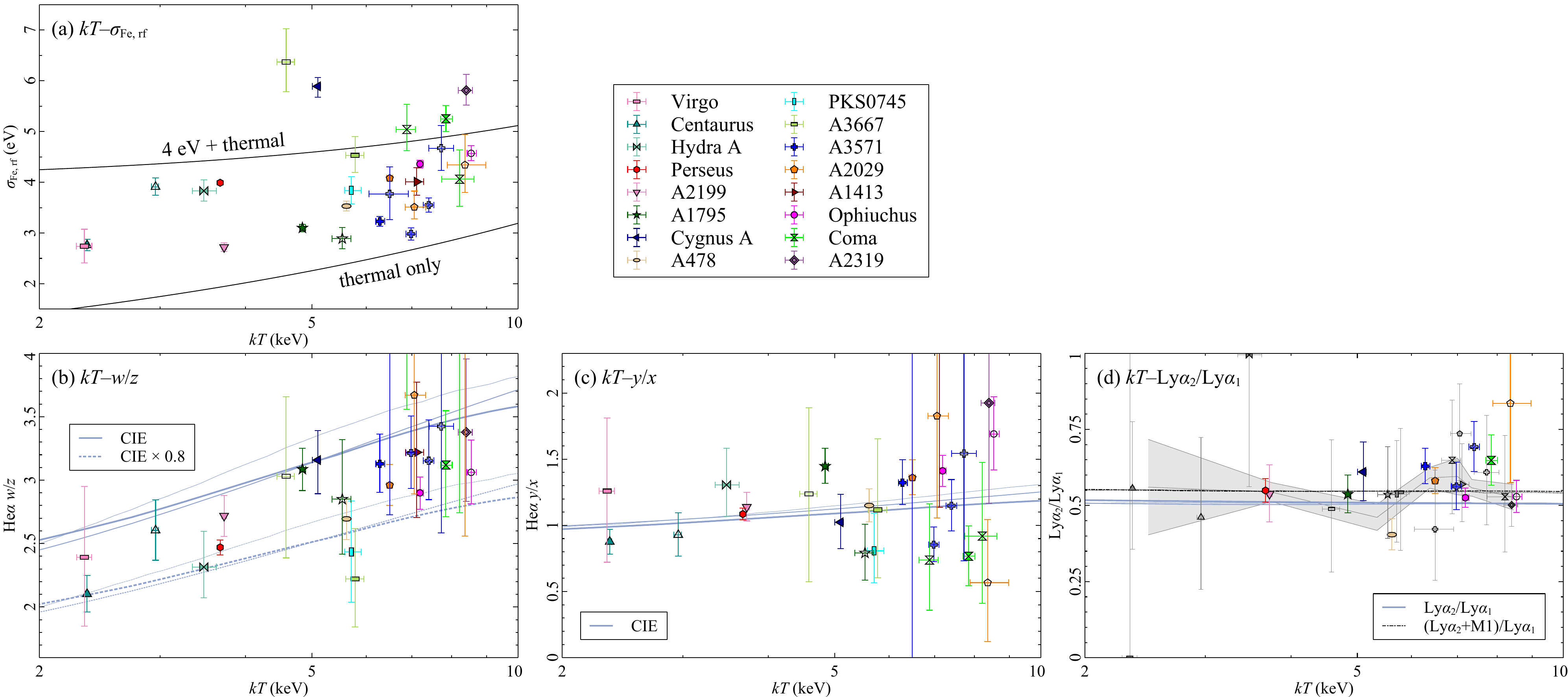}
\caption{(a) Relation between the rest-frame-corrected $\sigma_\textup{Fe}$ and $kT$.
Solid lines indicate the expected $\sigma_\textup{Fe}$--$kT$ relations for pure thermal broadening
and for thermal broadening with an additional non-thermal component of 4\,eV.
Different face-colour intensities indicate different fields within the same target,
as filled ones represent the central pointing.
(b, c, d) Line flux ratios of \hea{} $w$/$z$, \hea{} $y$/$x$, and Fe~\lyaa{} plotted against $kT$.
Thick, medium, and thin solid lines in each panel are CIE predictions
from \atomdb{}, \chianti{}, and \spex{}, respectively.
In the (b) panel, 80 per cent of the CIE prediction is also plotted in a dashed style.
In the (d) panel, results from the \lya{} doublet with fewer than 400\,counts are de-emphasised
by using thinner error bars, smaller markers, and a grey colour.
Four-point moving average is shown by the shaded area.
Dot-dashed line indicates the (Ly$\alpha_2$+M1)/Ly$\alpha_1$ ratio from \spex{}.
\label{fig:kt}}

\end{figure*}

\begin{figure}

\centering
\includegraphics[width=\columnwidth]{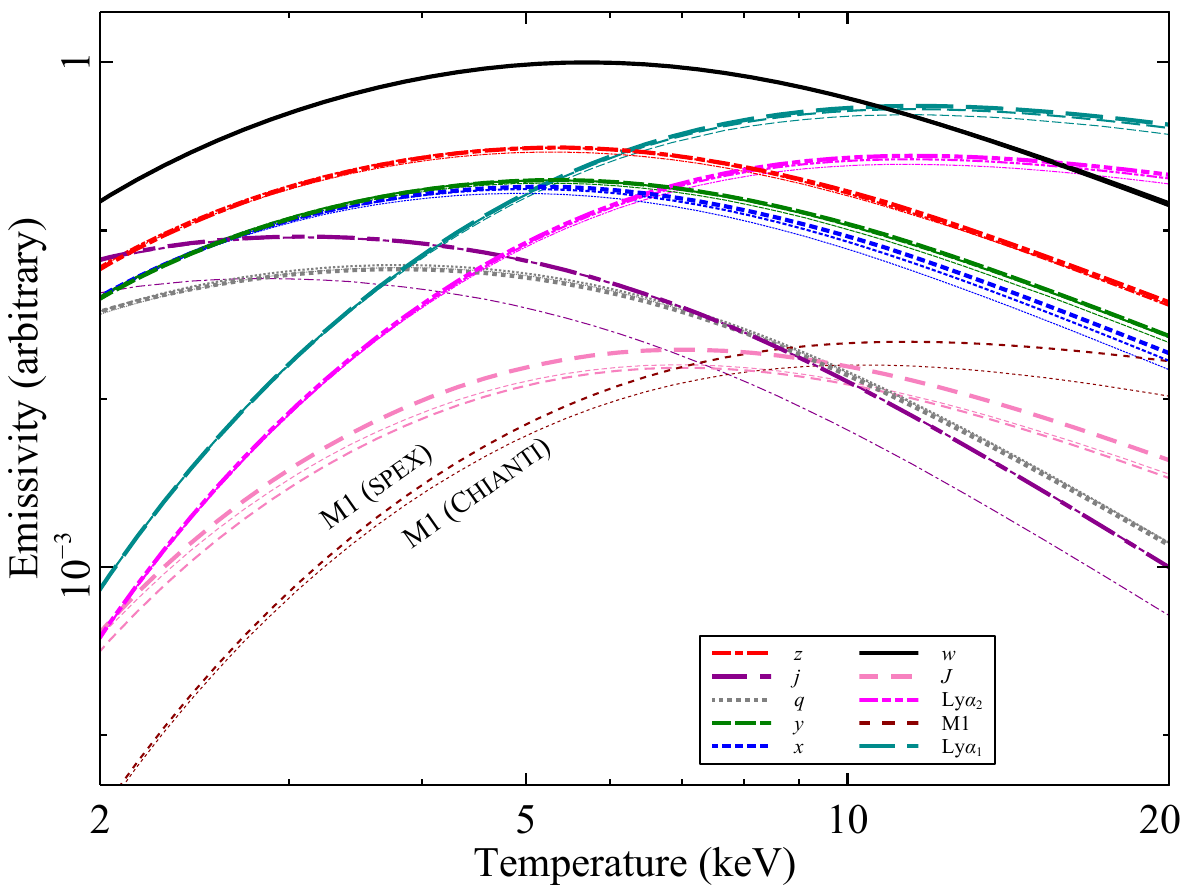}
\caption{Emissivities of lines listed in Table~\ref{tab:line}
derived from \atomdb{} version 3.1.3 (thick), \spex{} version 3.08.03 (medium), and \chianti{} version 11.0.2 (thin).
The ionisation fraction from \citet{Urdampilleta17} is adopted.
Values are normalised so that the peak emissivity of the \hea{}~$w$ line is unity.
\label{fig:emiss}}

\end{figure}

Table~\ref{tab:ratio} summarises the measured $kT$, $\sigma_\textup{Fe}$,
\lya{} intensity, and line flux ratios to the $z$ line.
The observed $kT$ values span a wide range, from 2\,keV in Virgo and Centaurus, which are in a classical cool-core regime,
to around 10\,keV in A2319, where heating due to a major merger is considered effective.

\subsection{$kT$ and $\sigma_\textup{Fe}$ \label{subsec:sigma}}

Figure~\ref{fig:kt}a shows the $kT$ dependence of measured line broadening
$\sigma_\textup{Fe}$ corrected to the rest frame for each object and field.
Since the $\sigma_\textup{Fe}$ value includes both effects due to thermal and turbulent motions,
the positive correlation between $\sigma_\textup{Fe}$ and $kT$ is naturally explained.
An additional point is that the observed line widths fall in the upper regime
of the CIE prediction, considering only the thermal broadening
$\sigma_\textup{Fe} = (E_\textup{Fe}/c) \sqrt{kT/m_\textup{Fe}}$
on the assumption that the ion temperature is the same as the electron temperature
\footnote{This assumption is reported to be reasonable in the ICM as far as two objects:
Perseus \citep[][]{Hitomi18a} and Centaurus \citep[][]{Kondo26}.
A counter-discussion for an extremely hot and disturbed case is predicted and summarised in \citet{Churazov26b}.}.
Thus, the measured $\sigma_\textup{Fe}$ values are substantially contributed to
by non-thermal motions with turbulence and unresolved bulk flows.
Most of our samples lie between the pure thermal-broadening case (see above)
and the case with an additional 4\,eV of broadening (about 180\,km\,s$^{-1}$ in the rest frame).
Our results, relying on the line-based measurement, are consistent with the picture
that many clusters exhibit subsonic turbulent velocities and a few per cent non-thermal pressures
within the Resolve array scale, excluding majestic merging systems
(e.g., A2319, \citealt{XRISM25h}; Coma, \citealt{XRISM25d})
and environments under strong feedback from the central engine \citep[e.g.,][]{Majumder26a}.
What the low non-thermal pressure supports in the ICM mean remains debated \citep[e.g.,][]{XRISM25f, McNamara26}
and is one of the top-priority questions to be addressed during the \xa{} era.

\subsection{\hea{} triplet \label{subsec:hea}}

\subsubsection{$w$/$z$ ratios \label{subsec:zw}}

\begin{figure}

\centering
\includegraphics[width=\columnwidth]{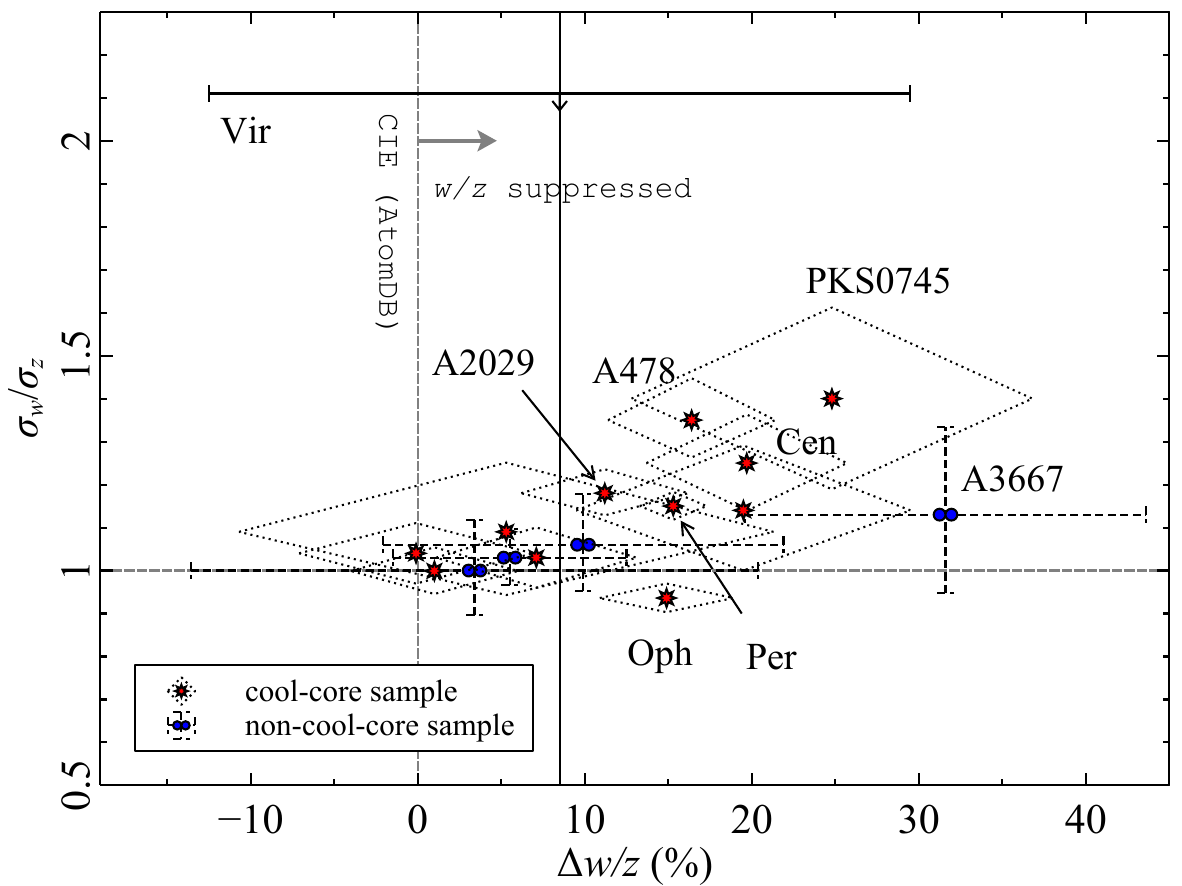}
\caption{Observed $\sigma_{w}$/$\sigma_{z}$ ratios plotted against the relative deviation of $w/z$.
Relative deviation is determined as
$\Delta w/z = (w/z|_\textup{CIE, \atomdb{}} - w/z|_\textup{obs})/w/z|_\textup{CIE, \atomdb{}}$;
thus, it is positive when the observed $w/z$ is suppressed.
The measured width is corrected for the natural broadening of each line,
though it is negligible ($1.3 \times 10^{-7}$\,eV for $z$ and 0.3\,eV for $w$ in FWHM).
\label{fig:zw}}

\end{figure}

We plot the \hea{} $w/z$ ratios against $kT$ in Fig.~\ref{fig:kt}b,
together with the CIE-predicted $w/z$--$kT$ relations calculated with
\atomdb{} version 3.1.3, \spex{} version 3.08.03 \citep[][]{Kaastra96, Kaastra26},
and \chianti{} version 11.0.2 \citep[][]{Dere97, Dufresne24}.
Predicted line ratios are derived from the emissivities of each line (Fig.~\ref{fig:emiss}).
The observed $w/z$ ratios generally follow the theoretical expectation
that hotter ICM plasmas exhibit larger $w/z$ values.
However, several clusters show systematically suppressed $w/z$ ratios relative to CIE predictions.
In particular, Centaurus, Perseus, A478, A2029, and Ophiuchus
show suppressions of approximately 20 per cent with significances of $\gtrsim 2\sigma$.
More moderate hints of low $w/z$ ratios are also observed in Hydra~A, A2199, and A3667,
though with larger uncertainties.
A well-established mechanism that reduces the $w/z$ ratio in the ICM is resonant scattering \citep[][]{Gilfanov87b}.
In this process, $w$-line photons are absorbed and re-emitted out of sight,
effectively suppressing the observed flux of optically thick lines in cluster centres \citep[e.g.,][]{Zhuravleva11}.
Because the optical depth of $w$ can be close to or even exceed unity in dense cluster cores,
the $w$ line is expected to be the most strongly affected transition among the lines listed in Table~\ref{tab:line}.

Several recent \xa{} studies of strong cool-core clusters have similarly reported suppressed $w/z$ ratios
and interpreted them as signatures of resonant scattering
(Centaurus, \citealt{Kondo26}; A2199, \citealt{Suda26}; PKS0745, \citealt{Tanaka26};
A2029, \citealt{XRISM25c}; Perseus, \citealt{Heinrich26b}).
An important characteristic of resonant scattering is the apparent broadening of the $w$ line.
Since resonant scattering preferentially suppresses the line core,
the observed $w$ profile becomes broader than those of optically thin lines
(see \citealt{Churazov10b} for a review, and \citealt{Hitomi18b, Kondo26} for applications to observational data).
Motivated by this expectation, we measure the widths of $w$ and $z$ separately
for the central brightest fields of each object.
In these fits, the width of the $w$ line ($\sigma_{w}$) is allowed to vary independently
from that of the other lines ($\sigma_{z}$), and all redshifts are tied.
Figure~\ref{fig:zw} shows the resulting line-width ratios, $\sigma_{w}/\sigma_{z}$, plotted against
the relative deviation $\Delta w/z = (w/z|_\textup{CIE, \atomdb{}} - w/z|_\textup{obs})/w/z|_\textup{CIE, \atomdb{}}$,
where $w/z|_\textup{obs}$ (and $kT$) is retrieved from the original values in Fig.~\ref{fig:kt}b.
Most cool-core clusters with suppressed $w/z$ ratios also exhibit broadened $w$ lines at $\gtrsim 2\sigma$ significance
while merging samples show consistent $\sigma_{w}$ and $\sigma_{z}$.
These results support the resonant-scattering interpretation for the cool-core systems in our sample.

Another explanation for the suppressed $w/z$ ratio is charge exchange.
In this process, electrons captured into highly excited states cascade towards the ground state,
thereby enhancing the $z$ line and reducing the $w/z$ ratio \citep[e.g.,][]{Gu18, GS23}.
Charge-exchange emission may become effective where hot ICM interacts with cold gas or dust,
as observed in starburst environments.
Alternatively, recombining NEI plasma may also enhance the $z$ line,
particularly in dynamically disturbed systems such as merging clusters.
In such cases, cascade emission can additionally strengthen higher-$n$ transitions
\citep[e.g., He$\gamma$, Ly$\delta$,][]{PD00}.
\citet{Chakraborty20b} also discuss that part of the fluxes of such higher-$n$ transitions, if optically thick,
can be redistributed to lower-energy transitions, resulting in an enhancement of $z$
(referred to as Case A-to-Case B transition).
In the two best targets to date, Perseus and Centaurus, extensive data from higher-$n$ transitions
suggest the presence of a multi-temperature CIE plasma \citep[][]{Hitomi18c, Kondo26},
despite a tentative hint of a charge-exchange contribution.
To investigate the charge-exchange or other origins, a more detailed discussion of interesting exceptions
in Fig.~\ref{fig:zw} will be important in the future:
A3667, showing highly suppressed $w/z$ by about 30 per cent even with a disturbed environment
(but an interface of cool and hot gas components may exist, \citealt{Omiya26a}),
and Ophiuchus with larger $\sigma_{z}$ than $\sigma_{w}$.

\subsubsection{$y$/$x$ ratios \label{subsec:xy}}

Figure~\ref{fig:kt}c shows the observed \hea{} $y/x$ ratios together with the CIE predictions.
Unlike the systematic $w/z$ suppression shown in Fig.~\ref{fig:kt}b,
the observed $y/x$ ratios do not exhibit a clear global trend.
The hotter systems, which generally show greater velocity broadening
and more dynamically disturbed environments (e.g., Coma and A2319),
have larger scatter and uncertainties in $y/x$,
likely because the intercombination lines are relatively weak.
Nevertheless, several clusters in the cooler and dynamically quieter regime,
including Centaurus, Hydra~A, A1795, A3571, A2029, and Ophiuchus,
show deviations from the CIE predictions at the level of 1--2$\sigma$.
Similar deviations have also been reported in previous \xa{} papers for some objects
(e.g., \citealt{Fujita25}, Ophiuchus; \citealt{XRISM25c}, A2029; \citealt{Sarkar26a}, A1795).

\citet{Sarkar26a} discussed uncertainty in atomic data for the $y$ line using A1795 observations.
They demonstrated that uncertainties in transition probabilities can significantly affect the predicted $y$ intensity,
since the $y$ line has one of the largest atomic uncertainties within the \hea{} complex \citep[][]{Hitomi18c, Chakraborty24}.
While theoretical uncertainties may play an important role, they may not fully explain the observed deviations,
as the measured ratios scatter both above and below the CIE predictions (Fig.~\ref{fig:kt}c).
The measured $y/x$ ratio can also be affected by nearby satellite lines,
such as the \ion{Fe}{xxiv} $q$ line (Table~\ref{tab:line}).
Because the emissivities of Li-like satellites peak at $kT \lesssim 4$\,keV (see Fig.~\ref{fig:emiss} for $j$ and $q$),
their contributions are expected to be more important in cooler systems, such as Centaurus and Hydra~A.

A spatially resolved study has reported that anomalous $y/x$ ratios do not necessarily
coincide with structures, such as cold fronts or radio relics \citep{Fujita25}.
\hirata{} are performing sub-array analyses of A3571,
currently one of the highest-quality and most densely mapped \xa{} cluster datasets.
Their preliminary results suggest a possible correlation between anomalous line ratios
and a temperature discontinuity of the ICM.
\citet{Suda26} provided a similar picture for A2199, with a strongly localised anomalous $y$/$x$ distribution.
If the observed deviations in $y/x$ have a physical origin, whether astrophysical or atomic,
they would require an enhanced population of levels that contribute to intercombination transitions.
Such behaviour may indicate cascade processes in addition to standard collisional excitation.

\subsection{\lya{} doublet \label{subsec:lya}}

\subsubsection{Unresolved contribution of the M1 transition \label{subsubsec:m1}}

We present the Fe~\lyaa{} ratios as a function of $kT$ in Fig.~\ref{fig:kt}d.
Both CIE models predict an almost $kT$-independent Fe~\lyaa{} ratio of about 0.50,
primarily determined by the statistical weights of the upper levels
($^2$P$_{1/2}$ and $^2$P$_{3/2}$).
Most samples are broadly consistent with this prediction.
Then, we focus on the objects and fields, excluding the de-emphasised measurements
with limited photon statistics (Fig.~\ref{fig:kt}d):
Perseus, A2199, A1795, Cygnus~A, A478, A3571, A2029, A3571N,
Ophiuchus, A3571S, Coma, A2029N2, and OphiuchusSW.
Fitting these 13 measurements with a constant value yields a ratio of $0.55 \pm 0.02$,
slightly higher than the predictions from \atomdb{}, \spex{}, and \chianti{}.
Excluding \textit{outliers} beyond the $1\sigma$ threshold from the CIE prediction
yields a consistent result of $0.54 \pm 0.02$.

The emission line from a magnetic-dipole transition from $^2$S$_{1/2}$ to $^1$S$_{1/2}$
(M1 line, Table~\ref{tab:line} and Fig.~\ref{fig:emiss}) can explain this baseline excess.
The \ion{Fe}{xxvi} M1 line is unresolvably coincident with Fe~Ly$\alpha_{2}$ \citep[e.g.,][]{Wong02, Yang25}.
Because \atomdb{} does not include this transition, both as a line and a pseudo-continuum,
our Gaussian component for Ly$\alpha_{2}$ intrinsically includes any unresolved M1 contribution.
The \spex{} and \chianti{} codes include the M1 line, adopting that the M1 flux is approximately
10 per cent of Ly$\alpha_{2}$ (Fig.~\ref{fig:emiss}).
\citet{Yang25} also provide a theoretical estimate of about a 10 per cent contribution of the unresolved M1 line.
Therefore, we assume $(\textup{Ly}\alpha_{2}+\textup{M1})/\textup{Ly}\alpha_{1} \sim 0.55$
independent of $kT$ (the dot-dashed line in Fig.~\ref{fig:kt}d),
and the observed baseline ratio is consistent with the contribution
from the M1 transition within the CIE framework.
The importance of M1 in radiative transfer codes, such as \textsc{cloudy},
has been confirmed in the interpretation of \xa{} observations \citep[][]{Gunasekera25b},
and this may also extend to the simpler case of the ICM.

\subsubsection{Possible origins of anomalous \lyaa{} \label{subsubsec:anom}}

A potentially more intriguing result is that hotter environments may preferentially exhibit larger Fe~\lyaa{} ratios.
Among the line ratios investigated in this work, Fe~\lyaa{} is the only quantity
whose deviation from the CIE prediction appears to vary with $kT$
(see the mean values in Fig.~\ref{fig:kt}d).
The first reports of anomalous \lya{} doublets in the ICM were presented
for A2029 \citep[][]{XRISM25c} and Coma \citep[][]{XRISM25d},
and several other systems show marginally large Fe~\lyaa{} ratios at the level of 1--2$\sigma$.
Furthermore, \citet{XRISM25g} identified similar structures in stacked spectra of 10 clusters,
including both \lya{} and Ni~Ly$\alpha$
\footnote{Even in the stacked spectrum, the Cr~Ly$\alpha$ doublet does not have sufficient photon statistics,
and the Ca~Ly$\alpha$ structures cannot be fully resolved.}.

At present, the origin of the plausible \lya{} anomaly remains unclear.
The likely $kT$ dependence is not statistically significant
because most of the mean values overlap
with the expected $(\textup{Ly}\alpha_{2}+\textup{M1})/\textup{Ly}\alpha_{1}$ ratio.
Therefore, we regard the current hint only as tentative.
Nevertheless, it is still useful to discuss several possible mechanisms
that may contribute to the observed behaviour,
excluding those requiring extreme physical environments,
such as a high-density plasma ($>10^{10}$\,cm$^{-3}$, \citealt{Kurihara26})
or a strong magnetic field ($\sim 10^7$\,G, a threshold of the Zeeman effect for \lya{}),
which are unlikely assumptions for ICM.

\begin{itemize}

\item Dielectronic recombination (DR) satellite lines can contribute to the observed Fe~\lyaa{} behaviour.
For the \ion{Fe}{xxvi} series, the DR satellites are resonantly excited
by electrons with energies of about 5\,keV \citep[e.g.,][]{Yang21}.
Considering both the Maxwellian electron distribution and Fe ionisation fraction,
rendez-vous of thermal electrons with 5-keV energy and H-like Fe ions in the CIE regime,
i.e., the chance of the \ion{Fe}{xxvi} DR processes, is peaked at $kT$\,$\sim$\,7\,keV.
Interestingly, the rolling average of the observed Fe~\lyaa{} ratios exhibits a marginal peak
around 6--7\,keV (Fig.~\ref{fig:kt}d).
Although the statistical significance is low, this coincidence
raises the possibility that unresolved DR satellites may contribute to the observed trend.

The contribution of DR satellites may be further enhanced if the electron distribution departs from a Maxwellian form.
Non-Maxwellian electrons are known to increase DR rates
and the corresponding satellite-line emissivities \citep[e.g.,][]{GP79}.
Similar anomalies in \lya{} (and \hea{}) structures have long been reported in solar-flare spectra
observed by the \textit{Hinotori} and \textit{Yohkoh} missions,
where DR satellites associated with a few keV non-thermal electrons \citep[e.g.,][]{Phillips08}
or high-energy suprathermal electrons \citep[e.g.,][]{Tanaka86, Pike96}
are discussed as a possible explanation.

In principle, the \ion{Fe}{xxv} $J$ line may provide an additional diagnostic
of non-Maxwellian distributions \citep[][]{Kaastra09}.
At present, however, few clusters possess sufficient photon statistics for such a measurement,
and the predicted emissivities of the $J$ (and similar \ion{Fe}{xxiv} DR $j$) line still differ
among atomic codes (Fig.~\ref{fig:emiss}).
Future high-statistics observations of the \ion{Fe}{xxiv}, \ion{Fe}{xxv}, and \ion{Fe}{xxvi} complexes,
together with improved laboratory measurements and atomic calculations of DR satellites,
will help assess the importance of these processes in the ICM.

\item Cascade-producing atomic processes may modify the populations
of specific excited states and thus alter the observed line emissivities.
For example, charge-exchange measurements performed with an electron beam ion trap (EBIT)
have demonstrated Fe~\lyaa{} ratios approaching unity (Fig.~23 of \citealt{Wargelin08}).
However, we note that the \hea{} complex is expected to be affected even more strongly,
resulting in measurable changes in the corresponding line ratios (see also Section~\ref{subsec:zw}).
In the present sample, we find no clear correlation between the deviations of
\hea{} $w/z$ and Fe~\lyaa{} from the CIE predictions (Figs.~\ref{fig:kt}b and \ref{fig:kt}d).

Another cascade-producing process is the recombination of highly charged ions.
Radiative recombination, followed by a de-excitation cascade, can redistribute electron populations
among the L-shell sublevels \citep[e.g.,][with anisotropic recombination]{Stoehlker97, Surzhykov02, Wong02, Wu07b}.
Such a process on bare Fe ions may modify the relative intensities of \lyaa{}, and the nearby M1 transition,
without impacting the line ratios within the \hea{} structures, unlike the CX process.
If recombination contributes substantially to the observed Fe~\lyaa{} ratios,
the tentative $kT$ dependence could, for example, partially reflect variations
in the fraction of bare Fe ions in the ICM.
The radiative recombination processes in CIE and NEI plasma
and possible contribution to the M1 enhancement will be discussed more in detail in \gu{}.

\item Given the possible asymmetric population of magnetic sublevels in $^2$P$_{3/2}$,
the Ly$\alpha_1$ emission can be intrinsically polarised,
unlike Ly$\alpha_2$ from $^2$P$_{1/2}$ and M1 from $^2$S$_{1/2}$.
Some EBIT experiments \citep[e.g.,][]{Robbins06, Bettadj10, Thorn11} have shown that
the observed \lyaa{} ratio strongly depends on this polarisation effect and the incident electron energy.
Crucially, polarising Ly$\alpha_1$ requires anisotropic electron collisions \citep[][]{GR21}.
On the other hand, the Debye length in typical intracluster environments is about
$2.35 \times (kT/\text{keV})^{1/2} (n_\text{e}/\text{cm}^{-3})^{-1/2}$\,km.
Since this length is extremely small compared to the scale of the Resolve FoV,
the ICM is effectively a quasi-neutral plasma.
If such polarisation effects are relevant in ICM observations,
they would imply the presence of anisotropic electron collisions associated with bulk plasma flows.
Magnetic fields and bulk velocities further complicate the geometry of these flows and collisions.
Observationally, the low Fe~\lyaa{} ratios reported in some clusters
(e.g., A478 in this work; A754 in \citealt{Omiya26b}) might support a contribution from such polarisation effects.
\end{itemize}

\subsection{Future prospects for diagnosing ICM plasma physics \label{subsec:pros}}

\begin{figure*}

\centering
\includegraphics[width=\textwidth]{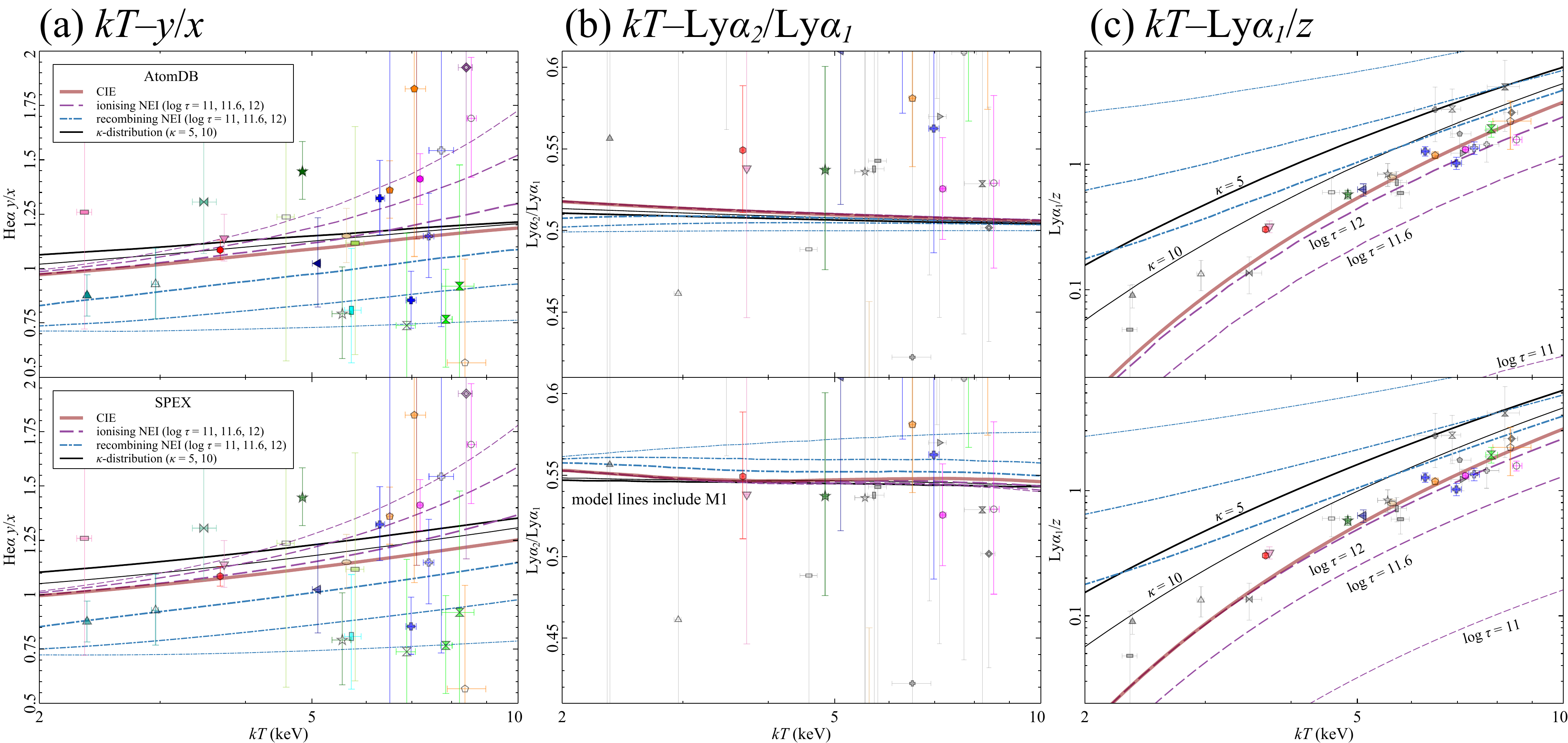}
\caption{
(a, b, c) Observed flux ratios of \hea{} $y/x$, Fe~\lyaa{}, and Ly$\alpha_{1}/z$, respectively, plotted against $kT$.
De-emphasised data points in (b) and (c) follow the same rule as Fig.~\ref{fig:kt}d.
Lines show the corresponding model predictions calculated with \atomdb{} (upper) and \spex{} (lower);
in panels (b), (Ly$\alpha_{2}$+M1)/Ly$\alpha_{1}$ is adopted.
Thick shaded lines represent the CIE predictions. Dashed and dash-dotted lines show ionising and recombining NEI plasmas,
respectively, for different ionisation timescales, with thinner lines corresponding to shorter timescales.
The thinnest lines correspond to $\tau=10^{11}$\,s\,cm$^{-3}$.
The ionising and recombining models assume initially neutral and fully stripped Fe, respectively.
Solid lines represent $\kappa$ distributions with $\kappa=5$ and $10$, where $kT$ is defined
as $(\kappa-1.5)/(\kappa+1)$ times the characteristic temperature of the corresponding $\kappa$ distribution.
\label{fig:toy}}

\end{figure*}

The mechanisms discussed above operate on wide physical scales,
from atomic transitions to macroscopic ICM motions.
As a first assessment of departures from the CIE framework,
we compare the observations with current model implementations
of ionising and recombining NEI plasmas over a range of ionisation timescales ($\tau$)
and of non-Maxwellian electrons described by $\kappa$ distributions
\citep[e.g.,][]{Meyer-Vernet95, LM09, HS15b} using \atomdb{} and \spex{}.
Figure~\ref{fig:toy}a demonstrates that ionising NEI plasmas can enhance $y$/$x$ in the hotter regime,
and recombining ones can reduce the ratio towards values observed in some anomalous samples.
The suprathermal electrons introduced by $\kappa$ also exhibit a slight enhancement of $y$/$x$.
In Fig.~\ref{fig:toy}b, the (Ly$\alpha_{2}$+M1)/Ly$\alpha_{1}$ ratio (\spex{}) is sensitive to
recombining NEI with low-ionisation ($\tau = 10^{11}$\,s\,cm$^{-3}$) achieving about 0.58
\footnote{As stated in Section~\ref{subsubsec:anom}, this effect will be further discussed in \gu{}.}
while other settings have less effect on the ratio.
Finally, Fig.~\ref{fig:toy}c shows that the observed Ly$\alpha_{1}$/$z$ values,
determined by the fraction of Fe$^{24+}$ and Fe$^{25+}$, are generally consistent with CIE.
Departures that can change $y$/$x$ and/or (Ly$\alpha_{2}$+M1)/Ly$\alpha_{1}$ drastically
also alter Ly$\alpha_{1}$/$z$ at the same time.
Thus, neither NEI cases nor $\kappa$ distributions in the current implementations
simultaneously explain the observed behaviour of the \hea{} and \lya{} diagnostics of anomalous samples.

The test result suggests that the observed deviations are unlikely to arise from a single modification
of the conventional plasma framework.
Numerical simulations can provide a useful next step, for example,
with the thermodynamic and ionisation histories of the ICM through shocks,
cold fronts, and other transient structures.
Forward modelling these simulations into spatially resolved Fe line ratios
and mock Resolve spectra would enable direct comparison
with the observables we demonstrated in this work,
particularly because the observed deviations may reflect conditions unique to individual objects or regions.
In Coma, the most unusual Fe~\lyaa{} ratio reaching unity is confined to a north-west sub-array region \citep[][]{XRISM25d}.
If each cluster or field is subject to various physical mechanisms as discussed here,
spatially resolved analysis (e.g., \hirata{} will showcase such a study for A3571)
and comparison with data from other wavebands will also gain importance.
An additional urgent and crucial observational examination involves expanding the sample of hotter galaxy clusters
over 10\,keV to investigate whether anomalies are uniformly observed in the hotter regime
or may exist within specified temperature ranges of the ICM.

The CIE paradigm continues to globally characterise the intracluster environment;
however, \xa{} observations have revealed notable deviations from the CIE assumption.
An enhanced comprehension of atomic physics will help to distinguish
possible processes discussed in this article from systematic uncertainties,
thereby refining our grasp of ICM physics.
Updating atomic databases and validating them against cosmic plasma
is essential to fully enjoy the high-resolution spectroscopic data from \xa{}
and to resolve fundamental questions in cluster science.
Our results highlight the critical role of interdisciplinary collaboration; for instance,
EBIT experiments specifically focusing on astrophysical plasma \citep[e.g.,][]{Amano26b, Hirata26}
are becoming increasingly significant in the era of \xa{} and subsequent missions.

\section{Conclusions \label{sec:concl}}
We have analysed the \xa{}/Resolve data of several clusters and measured the Fe line fluxes.
The results are summarised as follows, while the current preliminary results may lack statistical significance.

\begin{itemize}
\item[1.] The $w$/$z$ ratios well follow the predicted temperature dependence for most samples.
However, several cool-core clusters show substantially suppressed ratios by about 20 per cent
compared to the CIE expectations.
These samples also show more broadened $w$ lines than $z$,
which provides a hint of the resonant scattering effect in these cool cores.
On the other hand, disturbed samples have no evident suppression and/or
broadening of $w$, appearing to be less affected by the possible resonant scattering effect.

\item[2.] Although the $y$/$x$ ratios of some clusters diverge slightly from CIE predictions,
they are generally randomly distributed around the predictions, showing no correlation with ICM temperatures.
Since some studies hint at spatial variation of such a $y$/$x$ anomaly across cold fronts,
spatially resolved studies are the next important step.
It is also essential to accurately consider the relatively large atomic uncertainty intrinsic to the intercombination transitions,
along with the impact of satellite lines derived from low-ionised Fe species.

\item[3.] The observed Fe~\lyaa{} ratios are mostly consistent with the CIE value of 0.5,
and the mean value across samples is about 0.55, slightly higher than the prediction.
The unresolved M1 line contribution to the Ly$\alpha_{2}$ flux naturally explains this result.
Among the samples at 6--8\,keV of temperature, observed \lyaa{} values
even exceed the M1-corrected prediction.
While discussing some possible origins, we have not reached a robust conclusion at this stage.
Collaborative efforts of more samples covering a hotter regime, spatially resolved analyses,
and better understanding of atomic physics will be important to iron out the conundrum.
\end{itemize}

\begin{acknowledgements}
The authors appreciate the anonymous referee for their helpful comments and suggestions.
The authors also thank the \xa{} project team for leading mission development and science operations.
\xa{} has been developed and operated through an international collaborative effort
amongst the Japan Aerospace Exploration Agency (JAXA),
the National Aeronautics and Space Administration (NASA), and the European Space Agency (ESA).
In addition to the three space agencies, universities and research institutes from Japan,
the United States, and European states have joined to contribute to the development of the satellite,
scientific instruments, and data-processing software, and to further formulate scientific observation plans.
This research has made use of NED, which is operated by the Jet Propulsion Laboratory,
California Institute of Technology, under contract with NASA.
\chianti{} is a collaborative project involving George Mason University,
the University of Michigan (the United States), University of Cambridge (the United Kingdom),
and NASA Goddard Space Flight Center (the United States).
The figures in this report are created using \textsc{veusz} (\url{https://veusz.github.io/}).
YF acknowledges support from the Grants-in-Aid for Scientific Research (KAKENHI)
of the Japanese Society for the Promotion of Science (JSPS) grant Nos.~23H04899 and 25H00672.
HY acknowledges the auspices of JSPS KAKENHI grant Nos.~26H02075 and 23K25907,
and the National Institute for Fusion Science (NIFS) Collaboration Research Program NIFS25KRCQ001.
FM acknowledges financial support from the Centre National d'\'{E}tudes Spatiales (CNES),
France (ROR: \url{https://ror.org/04h1h0y33}) within the framework of the \xa{} mission.
KF acknowledges the use of \texttt{ChatGPT} (GPT-5.5; OpenAI) for the correction of \texttt{Python} typos
and language editing, as well as apologises for disastrously humanising their polite and sophisticated English sentences.
The \xa{} data sets analysed in this article are available or will be made public after one year of exclusive use
from the \xa{} data archive \footnote{\url{https://data.darts.isas.jaxa.jp/pub/xrism/data/obs/}}
\footnote{\url{https://heasarc.gsfc.nasa.gov/FTP/xrism/data/obs/}}.
\end{acknowledgements}

%
\bibliographystyle{mnras}
\bibliography{paper_ref}

\end{document}